\documentstyle[aps,prl,multicol,epsf]{revtex}

\begin{document}

\title{Evidence for coexistence of the superconducting gap and the pseudo -
gap in Bi-2212 from intrinsic tunneling spectroscopy}

\author{ V.M.Krasnov$^{a}$, A.Yurgens$^{b}$, D.Winkler$^{c}$,
P.Delsing and T.Claeson}

\address{Department of Microelectronics and Nanoscience,\\
 Chalmers University of Technology, S-41296 G\"oteborg, Sweden}

\date{\today }
\maketitle

\begin{abstract}

We present intrinsic tunneling spectroscopy measurements on small
Bi$_2$Sr$_2$CaCu$_2$O$_{8+x}$ mesas. The tunnel conductance curves
show both sharp peaks at the superconducting gap voltage and broad
humps representing the $c$-axis pseudo-gap. The superconducting gap
vanishes at $T_c$, while the pseudo-gap exists both above and
below $T_c$. Our observation implies that the superconducting and
pseudo-gaps represent different coexisting phenomena.

{PACS numbers: 74.25.-q, 74.50.+r, 74.72.Hs, 74.80.Dm}

\end{abstract}

\begin{multicols}{2}

The existence of a pseudo-gap (PG) in the quasiparticle density of
states (DOS) in the normal state of high-$T_c$ superconductors
(HTSC) has been revealed by different experimental techniques
\cite{Renner,Miyak,Suzuki,ARPES,Bianc,Oleg,Inter,Puchkov}. For a
review, see Refs.~\cite{Puchkov,Randeira,CDW,Halbritt}.
Up-to-date, there is no consensus about the origin of the PG, the
correlation between the superconducting gap (SG) and PG, or the
dependencies of both gaps on material and experimental parameters.
A clarification of these issues is certainly important for
understanding HTSC.

From surface tunneling experiments, it was concluded that the SG
is almost temperature independent~\cite{Renner,ARPES}. At $T>T_c$,
it continuously evolves into the PG, which can persist up to room
temperature. Furthermore, it was observed that such a
superconducting gap has no correlation with $T_c$ and continues to
increase in underdoped samples despite a reduction of
$T_c$~\cite{Renner,Miyak,ARPES}. This was the basis for a
suggestion that the PG-state at $T>T_c$ is a precursor of
superconductivity~\cite{Randeira}. On the other hand, surface
tunneling into HTSC has several drawbacks\cite{Mallet}, e.g. it is
sensitive to surface deterioration. The growing controversy
requires further studies with alternative techniques.

Intrinsic tunneling spectroscopy has become a powerful tool in
studying the quasiparticle DOS {\it inside} bulk single crystals
of layered HTSC~\cite{Suzuki,Inter,Schlenga,Gough,Winkler,Meso}
and thus avoiding the sensitivity to surface deterioration. First
experiments were recently attempted~\cite{Suzuki,Gough,Winkler} to
study the PG in mesas fabricated on surface of
Bi$_2$Sr$_2$CaCu$_2$O$_{8+x}$ (Bi-2212) single crystals.
Unfortunately, intrinsic tunneling experiments also have several
problems, such as internal heating and stacking faults (defects)
in the mesas. To reduce overheating, a pulse technique was applied
by Suzuki {\it et al},~\cite{Suzuki}. The result at $T\sim T_c$
was essentially similar to the surface measurements~\cite{Renner},
leaving the obscure relationship between SG and
PG~\cite{Suzuki,Gough} unresolved.

In this paper we present results of intrinsic tunneling
spectroscopy for Bi-2212 mesas with considerably smaller areas,
compared to previous studies~\cite{Suzuki,Gough}. Smaller areas
allowed us to avoid stacking faults in the mesas and to avoid
mixing between the $c$- axis and $ab$- plane transport. As a
result, clean and clear tunnel-type current-voltage (I-V)
characteristics were observed, which allowed us to distinguish
superconducting and pseudo - gaps in a wide range of temperatures.
In contrast to surface and earlier intrinsic tunneling
experiments~\cite{Renner,Suzuki,Gough}, we have clearly traced
different behaviors of the SG and PG. Thorough studies of I-V
curves close to $T_c$ revealed that the superconducting gap does
vanish, while the PG does not change at $T=T_c$. All this speaks
in favor of different origins of the two coexisting phenomena and
against the precursor-superconductivity scenario of the PG.
Finally, we discuss interplay between Coulomb interaction and low
dimensionality as a possible mechanism for the c-axis PG in an
inherent two-dimensional (2D) system, such as the Bi-2212 single
crystals.

Mesas with different dimensions from 2 to 20 $\mu $m were
fabricated simultaneously on top of Bi-2212 single crystals. To
reduce the mesa area we adopted a self-alignment technique, see
Ref. ~\cite{Meso} for details of sample fabrication. The {\it
c}-axis I-V characteristics were measured in a three-probe
configuration. The contact resistance was small, about two orders
of magnitude less than the total resistance of the mesa at the
corresponding current. All the leads to a mesa were filtered from
high-frequency electrical noise. Altogether, more than 50 mesas
made on different crystals were investigated. Parameters of the
mesas are listed in Table 1. The three figures in the mesa number
represent the batch number, the crystal number and the number of
the mesa on the crystal, respectively. Letters "Ar" or "Ch"
indicate whether the mesa was made by Ar-ion or chemical etching.
Here we present results for slightly overdoped ($T_c=89$~K) and
optimally doped ($T_c=93-94$~K) samples. The pristine crystals
were slightly overdoped. Overdoped mesas were obtained by wet
chemical etching, which does not significantly change the oxygen
content. Optimally doped mesas were made by Ar-ion etching, during
which mesas partly loose oxygen. Such mesas had larger $T_c$,
$c$-axis resistivity, $\rho_c$ see Table 1, and pseudo-gap, see
Fig. 4.

Our fabrication procedure provides samples with highly
reproducible properties. This is illustrated in the

\begin{figure}
\noindent
\begin{minipage}{0.48\textwidth}
\epsfxsize=0.9\hsize \centerline{ \epsfbox{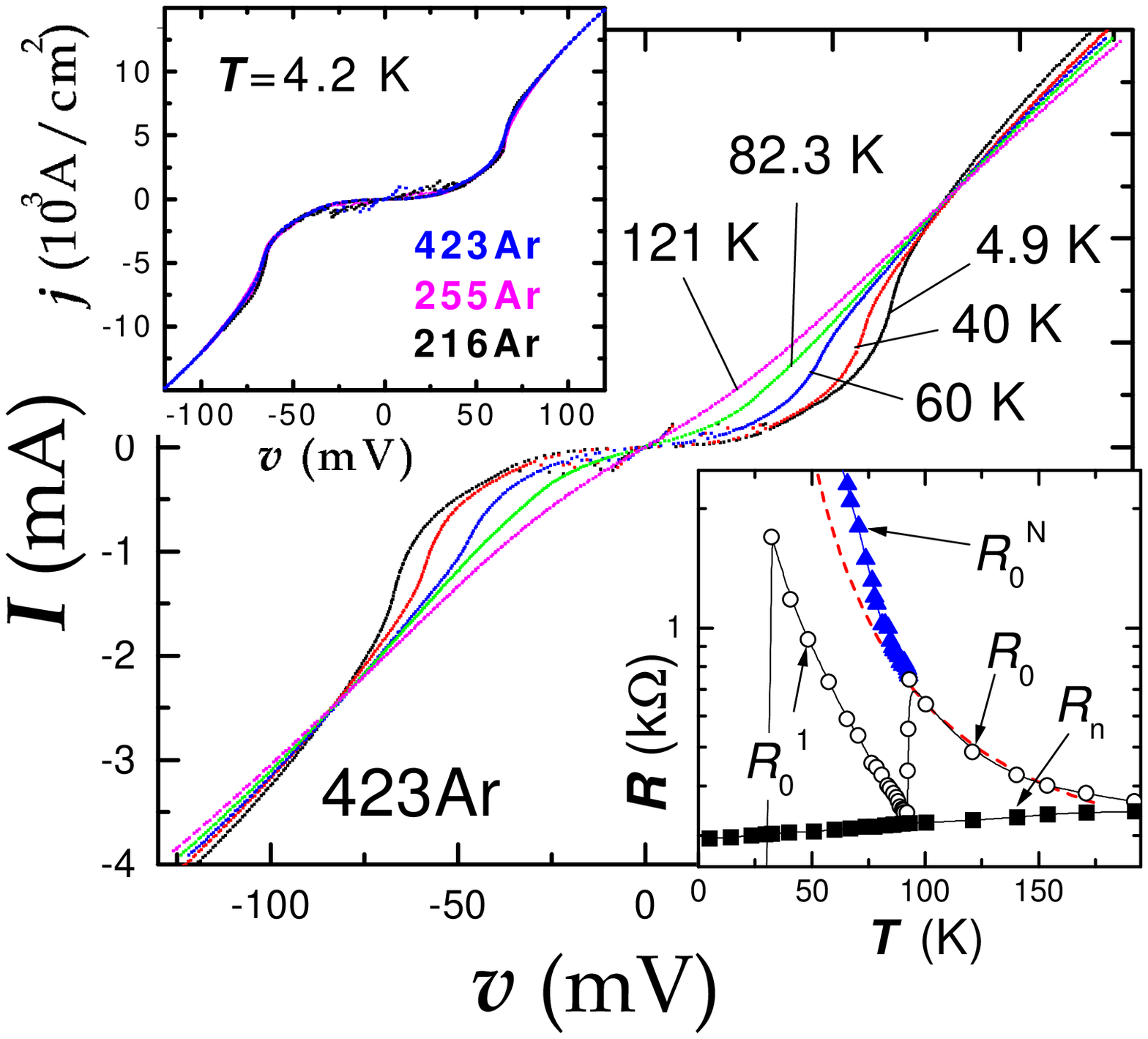} }
\vspace*{6pt} \caption{I-V curves per junction for the 423Ar mesa
at different $T$. Top inset shows normalized I-V curves at
$T$=4.2K for three different mesas. Bottom inset shows the
temperature dependence of the zero bias resistance, $R_0$ (open
circles), large bias resistance, $R_n$ (solid squares) and the
total sub-gap resistance, $R_0^N$ (solid triangles).}
\end{minipage}
\end{figure}

\noindent top inset of Fig.~1, in which current density, $j=I/S$,
vs. voltage per junction, $v=V/N$, curves at $T=4.2$~K are shown
for three mesas with different areas from different batches and
crystals. Here $S$ is the area and $N$ is the number of intrinsic
Josephson junctions (IJJ's) in the mesa. We note, that all
normalized I-V curves collapse into a single curve. In the
following, we will denote quantities corresponding to the whole
mesa by capital letters, and those related to an individual IJJ,
by small letters. The subscripts "s", "pg" and "n" will correspond
to the {\underline s}uperconducting, {\underline
p}seudo-{\underline g}ap and {\underline n}ormal state properties,
respectively.

In Fig. 1, $I-v$ curves and in Fig. 2 the voltage dependence of
the dynamic conductance $\sigma(v)= {\rm d}I/{\rm d}v(v)$ are
shown for the optimally doped mesa 423Ar at different
temperatures. Figs. 1 and 2 exhibit a typical tunnel-junction
behavior. At large bias current, there is a well defined
normal-state part of tunneling I-V curves with tunnel resistance
$R_n$. $R_n$ decreases by merely $\sim$ 15\% from 300 K to 4.2 K
and has no feature at $T=T_c$, as shown in the bottom inset of
Fig. 1. This is in accordance with the pure tunnel junction
behavior, for which $R_n$ is expected to be temperature
independent. The weak $T$-dependence of $R_n$ indicates an absence
of mixing between $c$-axis and $ab$-plane transport in our mesas.
Previously, however, a strong change of $R_n$ at $T=T_c$ has been
reported for larger mesas~\cite{Suzuki,Gough}. On the other hand,
the zero bias resistance, $R_0$, has a strong temperature
dependence, see bottom inset in Fig. 1. Below $T_c$, $R_0$ is
determined by the sub-gap resistance of the first IJJ, $R_0^1$, At
$T<$40K, a small critical current in the first IJJ appears, see
Fig. 3 a), and $R_0$ drops to the contact

\begin{figure}
\noindent
\begin{minipage}{0.48\textwidth}
\epsfxsize=0.9\hsize \centerline{ \epsfbox{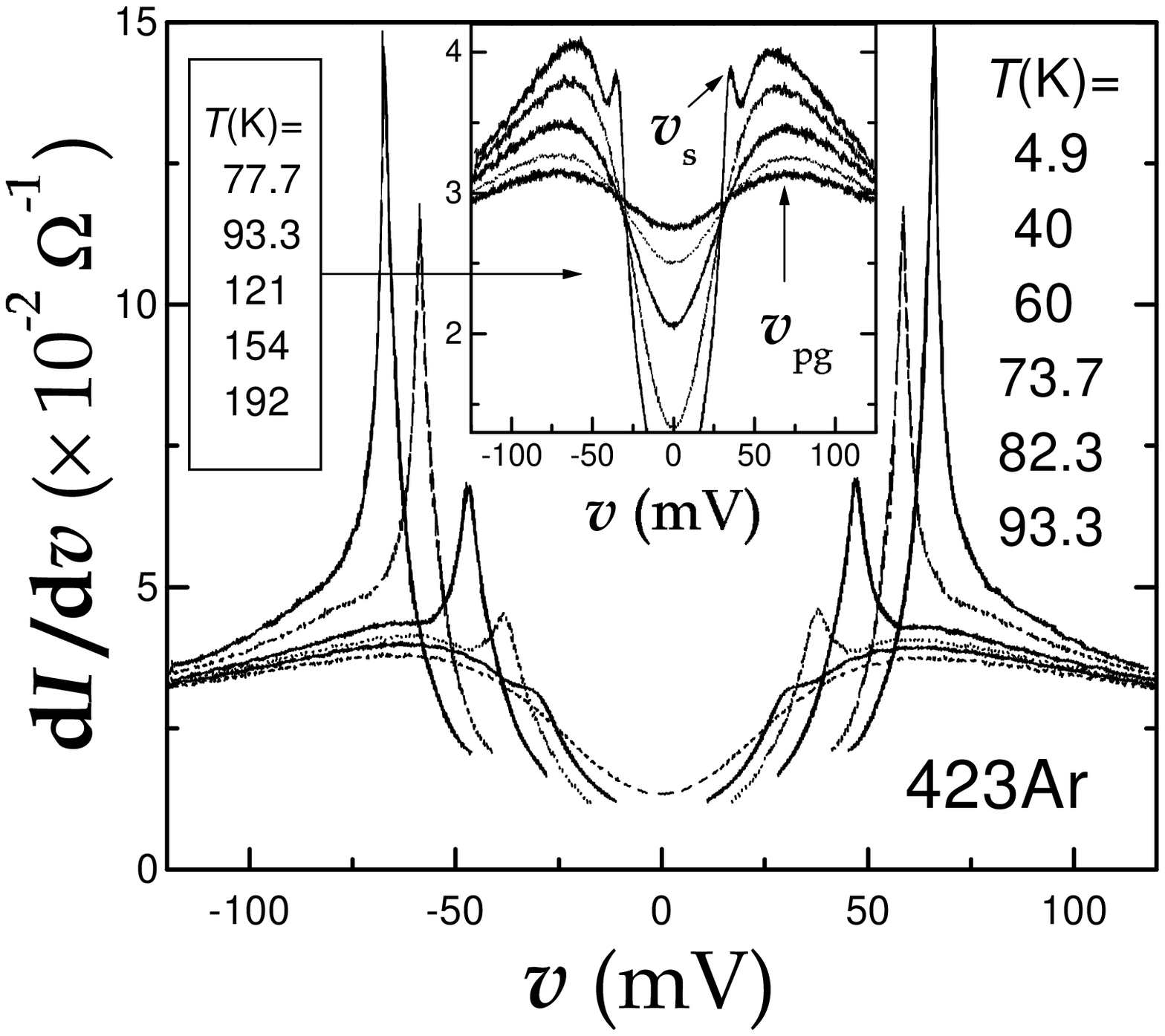} }
\vspace*{6pt} \caption{Dynamic conductance, $\sigma(v)$, at
different temperatures for 423Ar mesa. Inset shows detailed curves
for high $T$. Coexistence of the superconducting peak, $v_s$, and
the pseudo-gap hump, $v_{pg}$, is clearly visible at $T=77.7$ K. }
\end{minipage}
\end{figure}

\noindent resistance. Such a two stage decrease of $R_0$ is due to
a deterioration of IJJ's at the surface of the mesa \cite{Lee}.

At low $T$, there is a sharp peak in $\sigma(v)$, which we
attribute to the superconducting gap voltage, $v_s=2\Delta_s /e$.
With increasing $T$, the peak at $v_s$ reduces in amplitude and
shifts to lower voltages, reflecting the decrease in
$\Delta_s(T)$. At $T\sim 83\ {\rm K}$ $(< T_c \simeq 93$~K), the
superconducting peak is smeared out completely and only a smooth
depletion of $\sigma(0)$ (a dip) plus a hump in conductance at
$v=v_{pg}\simeq 70$~mV remain. The dip and the hump are correlated
to each other and both flatten simultaneously with increasing $T$,
see inset in Fig. 2. Therefore, both reflect the existence of the
pseudo-gap in the tunneling DOS. The $\sigma(0)$ gradually
increases with temperature but the I-V curves remain non-linear
nearly up to room temperature. At $T>T_c$, the zero-bias
resistance, $R_0$, is fairly well described by the
thermal-activation formula,

\begin{equation}
R_0 \propto exp( T^* /T), \ T^* \simeq 150 \pm 20 K,  \label{Eq1}
\end{equation}

\noindent as shown by the dashed line in bottom inset of Fig. 1.

In agreement with surface tunneling
experiments\cite{Renner,Miyak}, there are no sharp changes at
$T_c$. As shown in bottom inset of Fig. 1, at $T<T_c$, $R_0$
evolves continuously into the total (all $N$ junctions in the
resistive state) sub-gap resistance, $R_0^N$. This implies that
the PG persists also in the superconducting state. The gradual
evolution of the PG hump upon cooling through $T_c$ is most
clearly shown in inset of Fig. 2. It is seen that the PG dip/hump
feature does not change qualitatively upon cooling through $T_c$.
Moreover, the I-V curve at $T=$77.7 K shows that the
superconducting peak at $v_s$ emerges on top of the PG - features
which demonstrates a coexistence of both SG and PG features. From
Fig. 2 it is seen that by further decreasing the temperature, the
superconducting peak

\begin{figure}
\noindent
\begin{minipage}{0.47\textwidth}
\epsfxsize=0.9\hsize \centerline{ \epsfbox{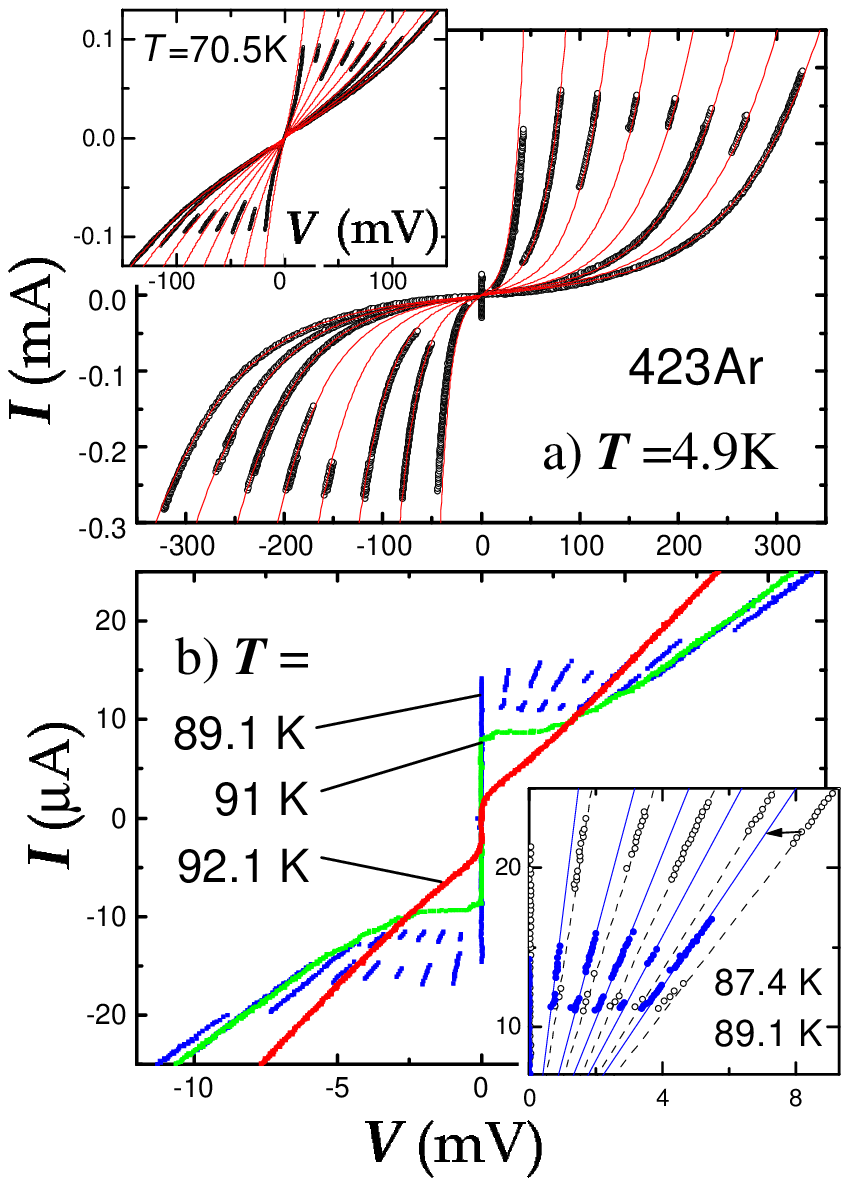} }
\vspace*{6pt} \caption{Detailed views of QP branches for the 423Ar
mesa at different temperatures. Thin lines in Fig. a) represent
polynomial fits and indicate good scaling of QP branches. The
inset in b) shows the first five QP branches at $T$ close to
$T_c$, in an expanded scale. The arrow demonstrates the tendency
for vanishing $\delta v_s(T)$ at $T\rightarrow T_c$ even when
measured at one and the same current. }
\end{minipage}
\end{figure}

\noindent shifts to higher voltages, increases in amplitude and
eventually the PG hump is washed out by the much stronger
superconducting peak. For optimally doped mesas, the PG hump can
be resolved at $T>$60 K, i.e. well below $T_c$. The gradual
opening of $\Delta_s$ at $T<T_c$, in addition to the PG, can also
be seen from a steeper growth of the total sub-gap resistance,
$R_0^N$, at $T<T_c$, as compared to the thermal-activation
behavior of $R_0$ at $T \ge T_c$, as shown in bottom inset in
Fig.1.

At low bias and $T<T_c$, multiple quasiparticle (QP) branches are
seen in the I-V curves, representing a one-by-one switching of the
IJJ's into the resistive state~\cite{Suzuki,Schlenga}. A detailed
view of multiple QP branches is shown in Fig. 3 for different $T$.
Dots and thin lines in Fig. 3 a) represent the experimental points
and a polynomial fit, correspondingly. Only the last branch,
having many data-points, was actually fitted, all the other thin
lines were obtained by dividing the voltages of this fit, $V_{\rm
fit}(I)$, by the integer number $N^*=N-n+1$, where $n$ is the
number of IJJ's in the resistive state. A good scaling of QP
branches is seen, which implies that there is no significant
overheating of the mesa at the operational current. If there were
overheating, $V_{\rm fit}(I)/N^*$ would not go

\begin{figure}
\noindent
\begin{minipage}{0.48\textwidth}
\epsfxsize=0.9\hsize \centerline{ \epsfbox{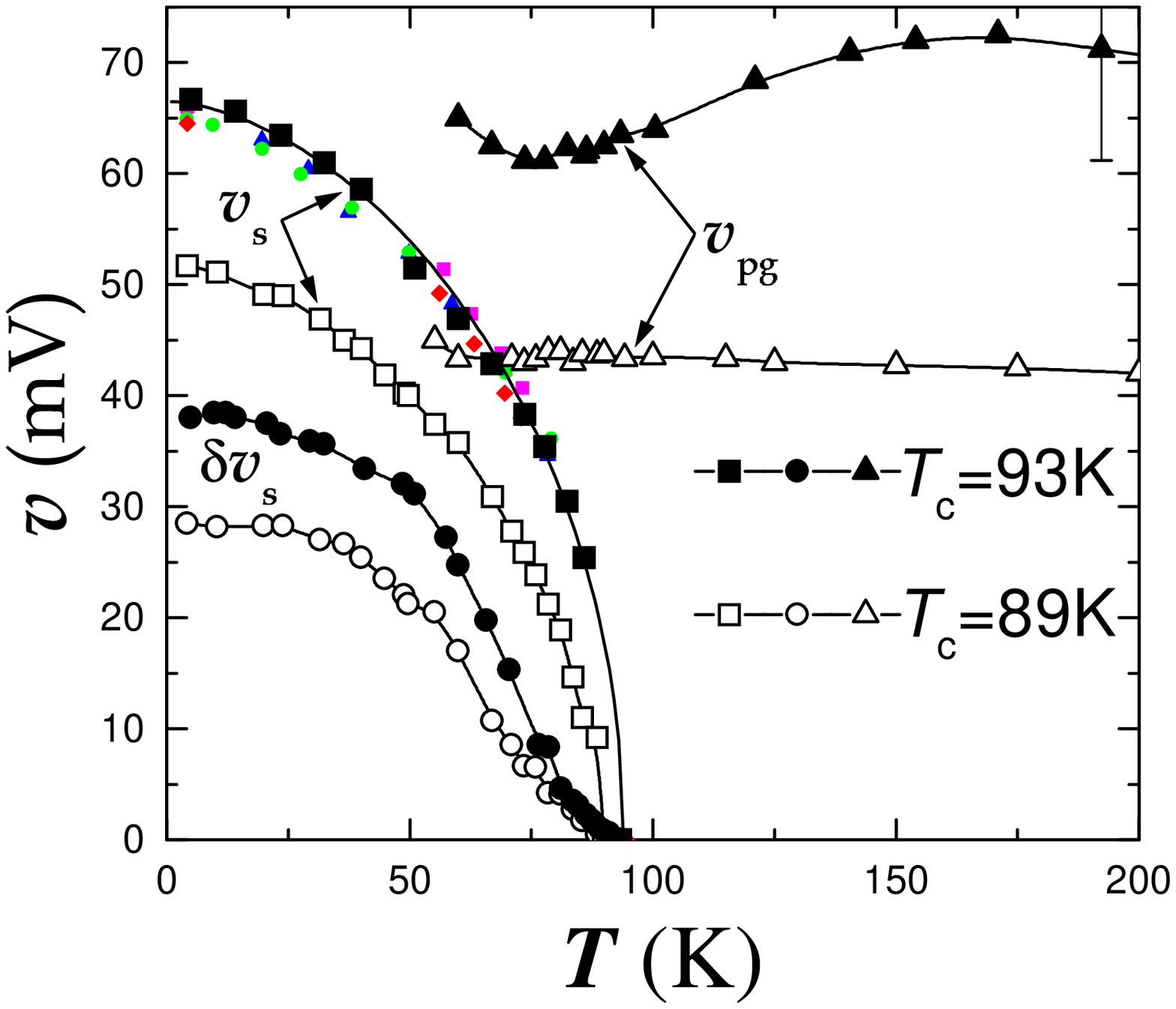} }
\vspace*{6pt} \caption{Temperature dependence of parameters of the
optimally doped (solid symbols) and overdoped (open symbols)
samples: the superconducting peak voltage, $v_s=2\Delta_s/e$, the
spacing between QP branches, $\delta v_s$, and the pseudo-gap hump
voltage, $v_{pg}$. It is seen that the superconducting gap
vanishes at $T_c$, while the pseudo-gap exists both above and
below $T_c$.}
\end{minipage}
\end{figure}

\noindent through the data points because switching of additional
IJJ's would cause a progressive increase of the internal
temperature and the branches with increasing count numbers would
have lower voltages due to the strong temperature dependence of
$R_0^N$ and $\Delta_s$.

The separation between QP branches, $\delta v_s$, is the
additional quantity, provided by intrinsic tunneling spectroscopy,
which can be used to estimate $\Delta_s$ in a wider range of
temperatures. From Fig.~3 b) it is seen that multiple QP branches
are clearly distinguishable up to $T\sim T_c-2$~K. From Table~1 it
is seen that $\delta v_s$ scales with $\Delta_s$. The $\delta v_s$
is less than $V_s/N$ simply because the critical current, $I_c$,
is less than $V_s/R_n$ and all IJJ's switch to the resistive state
before they reach the gap voltage, see Fig.~1. The $\delta v_s
(I=I_c)$ continuously decreases with $T$ and vanishes at $T_c$. In
principle, the temperature dependence of $I_c$ is also involved in
$\delta v_s(T)$, since we measure $\delta v_s$ at $I\approx I_c$.
However, the inset in Fig.3 b) reveals that $\delta v_s(T)$ still
tends to vanish at $T\rightarrow T_c$ even if we evaluate $\delta
v_s$ at one and the same current for all $T$.

In Fig.~4, the temperature dependencies of the superconducting
peaks, $v_s$ (squares), $\delta v_s (I=I_c)$ (circles), and
pseudo-gap humps, $v_{pg}$ (triangles), are shown for optimally
doped (solid) and overdoped (open symbols) samples. Small solid
symbols represent $v_s$ for the rest of the mesas listed in
Table~1, and the lines are guides for the eye. In agreement with
previous studies, both $v_s$ and $v_{pg}$ increase upon going from
overdoped to optimally doped samples\cite{Renner,Miyak,ARPES}. The
superconducting gap deduced from the sum-gap voltage
$V_s=2N\Delta_s/e$ is $\Delta_s(4.2K)\simeq 33$~meV for the
optimally doped sample, and $\simeq 26$~meV for the overdoped one.
In contrast to
surface tunneling experiments, we observe that $\Delta_s$
decreases considerably with temperature. The robust decrease of
$\Delta_s(T)$ from 4.2K to $T_c$ is more than 80 \% for the
overdoped mesas. Moreover, we can measure $\delta v_s (I=I_c)$ in
a wider range of $T$ and observe that it vanishes at $T\rightarrow
T_c$.

All this brings us to the conclusion that the superconducting gap
does close at $T_c$, in agreement with the previous observations
of vanishing of the superfluid density (divergence of the magnetic
penetration depth)~\cite{Lambda} and the Josephson plasma
frequency~\cite{Jplasma}. On the contrary, the PG is almost
temperature independent and exists both above and below $T_c$.
Therefore, the SG is not developing from the PG, and these two gaps
represent different coexisting phenomena. The recently observed
independence of the PG on magnetic field~\cite{Oleg} supports our conclusion
and also casts doubts about the precursor-superconductivity origin of the PG~\cite{Randeira}.

One possible "non-superconducting" PG-scenario is the formation of
charge or spin density waves (CDW or SDW)~\cite{Bianc,CDW}. HTSC's
are composed of quasi-two dimensional electronic systems with a
certain degree of Fermi-surface nesting~\cite{Bianc}, which can
make the system unstable with respect to CDW or SDW
formation~\cite{Gruner,Nesting}. A CDW or SDW is accompanied by a
PG in DOS, detectable by a surface-tunneling
spectroscopy~\cite{Inger}. Many similarities exist between the PG
in CDW or SDW (including ARPES~\cite{Nesting}, optical
conductivity and NMR~\cite{Gruner,CDW}) and the PG in HTSC. On the
other hand, an opening of the gap due to CDW or SDW is typically
accompanied by a metal-insulator transition~\cite{Gruner}, while
the $ab$-plane resistivity in Bi-2212 shows the opposite tendency
~\cite{Watanabe}.

We would also like to emphasize a similarity between the PG
features of $c$-axis tunneling in HTSC and Coulomb PG for
tunneling into a two-dimensional electron system (2DES). The
Coulomb PG in 2DES is well studied in connection with
semiconducting heterostructures~\cite{Chan,Efros,Levitov}.
Experimental $\sigma (v)$ curves from the inset in Fig.~2 are
strikingly similar to "V-shaped" tunneling characteristics of
2DES~\cite{Chan,Efros}. Certainly, the electron system in Bi-2212
is highly two-dimensional. Moreover, a Coulomb origin of the HTSC
pseudo-gap would naturally explain the increase of PG with
decreasing O-doping and carrier concentration. A large Coulomb PG
in low conducting 2DES is due to unscreened long-range Coulomb
interaction~\cite{Efros} and/or slow charge accommodation
\cite{Levitov}. Large PG could also appear if tunneling occurs via
intermediate low conducting BiO layers \cite{Halbritt}.

An attractive feature of both CDW/SDW and Coulomb PG scenaria is
that the PG can persist in the superconducting state. Below $T_c$,
SG and PG are combined into a larger overall gap\cite{CDW}. This
is in agreement with a definite trend for the increase of $v_{pg}$
at $T<T_c$, see Fig. 4. This might also help in understanding of
large "superconducting" gaps seen in underdoped HTSC
\cite{Renner,Miyak}. Whether the CDW/SDW or Coulomb PG scenaria
can explain all PG features in HTSC remains to be clarified.

In Conclusion, small mesa structures were used for intrinsic
tunneling spectroscopy of Bi-2212. We were able to distinguish and
simultaneously observe both superconducting and pseudo gaps in a
wide range of temperatures. The superconducting gap has a strong
temperature dependence and vanishes at $T_c$, while the pseudo-gap
is almost temperature independent and exists both above and below
$T_c$. This suggests that the pseudo-gap is not directly related
to superconductivity.

\begin{table}
\noindent
\begin{minipage}{0.47\textwidth}
\caption{Parameters of Bi2212 mesas, where $\rho_c$ is the
$c$-axis normal-state resistivity at large bias current.}
\begin{tabular}{ccccccc}
mesa & $S$ & $N$ & $T_c$ & $\Delta_s$(0) & $\delta v_s$(0) &
$\rho_c$(4.2K) \\ \space &($\mu$m$^2$) & \space & (K) & (meV) &
(mV) & ($\Omega$cm)\\ \hline
423Ar & $3.5\times 7.5$ & 10 & 93 & 33.3 & 38.5 & 44.9\\
255Ar & $5.5\times 6$ & 12 & 92.5 & 32.5 & 35.5 & 45.4\\
251Ar & $6\times 6$ & 12  & 92.5 & 32.5 & 35.5 & 44.5\\
211Ar & $4\times 7.5$ & 12 & 94 & 33.0 & 37 & 44.0\\
216Ar & $4\times 20$ & 10 & 94 & 32.3 & 38.5 & 44.9\\
015Ch & $12\times 15$ & 9 & 89 & 25.8 & 28.5 & 32.3
\end{tabular}
\end{minipage}
\end{table}

\end{multicols}

\end{document}